\documentclass[12pt]{article}
\usepackage{amssymb}

\begin{document}

\bigskip\ 

\begin{center}
\textbf{RELATIVISTIC TOP DEVIATION EQUATION }

\textbf{AND GRAVITATIONAL WAVES}

\bigskip\ 

\smallskip\ 

J. A. Nieto$^{\ast }$\footnote{%
nieto@uas.uasnet.mx}, J. Saucedo$^{\ddagger }$\footnote{%
jsaucedo@cajeme.cifus.uson.mx} and V. M. Villanueva$^{\dagger }$\footnote{%
vvillanueva@ifm1.ifm.umich.mx}

\smallskip\ 

$^{\ast \ddagger }$Departamento\textit{\ de Investigaci\'{o}n en F\'{i}sica
de la Universidad de Sonora,}

\textit{83000 Hermosillo Sonora, M\'{e}xico}

$^{\ast }$\textit{Facultad de Ciencias F\'{i}sico-Matem\'{a}ticas de la
Universidad Aut\'{o}noma}

\textit{de Sinaloa, 80010 Culiac\'{a}n Sinaloa, M\'{e}xico}

$^{\dagger }$\textit{Instituto de F\'{i}sica y Matem\'{a}ticas de la
Universidad Michoacana de San Nicol\'{a}s de Hidalgo, Morelia,
Michoac\'{a}n, M\'{e}xico, P. O. Box 2-82}

\textit{\ }

\bigskip\ 

\textbf{Abstract}
\end{center}

By using the relativistic top theory, we derive a relativistic top deviation
equation. This equation turns out to be a generalization of the geodesic
deviation equation for a pair of nearby point particles. In fact, we show
that when the spin angular momentum tensor associated to the top vanishes,
such a relativistic top deviation equation reduces to the geodesic deviation
equation for spinless point particles. Just as the geodesic deviation
equation for spinless particles can be used to investigate the detection of
gravitational waves, our generalized formula for a relativistic top can be
used to study the gravitational wave background. Our formulation may be of
special interest to detect the inflationary gravitational waves via the
polarization of the cosmic background radiation.

\bigskip\ 

Pacs numbers: 04.60.-m, 04.65.+e, 11.15.-q, 11.30.Ly


March, 2003

\newpage\ 

\noindent \textbf{I. INTRODUCTION}

\bigskip

It is well recognized that the geodesic deviation equation (GDE) for
spinless particles plays an important role in the search of gravitational
wave detectors [1-2]. Indeed, it is known that all of the projects currently
used to detect gravitational waves, including LIGO [3], VIRGO [4] and LISA
[5], have among their root physical bases such an equation. However, some
years ago, in references [6] and [7], it was proposed to use the
relativistic top equations of motion (RTEM) [8] (see also Refs. [9] and
[10], and references there in) instead of the GDE for the same purpose. The
main motivation for this alternative emerged from the observation that the
motion of a relativistic top is influenced by a gravitational force
involving the Riemann tensor just in a similar way as two spinless particles
in the GDE involve the Riemann tensor. In Ref. [6] general solutions of the
RTEM for the case of a top interacting with a gravitational wave were
investigated, while in Ref. [7] such solutions were applied to the specific
case of considering pulsars as gravitational wave detectors. Although the
alternative method proposed in Refs. [6] and [7] to detect gravitational
waves is interesting by itself, it has to be extended with the purpose of
comparing, in a direct way, the results derived from the RTEM with those
obtained \ from the method based on the GDE. Hence, instead of treating the
two alternatives as two independent methods, in this work we combine them
and we obtain a generalization of the GDE formalism. Specifically, we derive
a relativistic top deviation equation (RTDE) which is reduced to the GDE
when the spin tensor associated to the top vanishes.

It is known that binary pulsar systems may be important sources of
gravitational waves [11]. Here, as an application of the RTDE we shall argue
that binary pulsars can also be used as detectors of gravitational waves.
For this effect to be viable it is necessary that the binary pulsar system
has a companion source of gravitational waves. A similar but not quite the
same idea has already been considered by Laguna and Welszczan [12]. These
authors consider a rotating black hole as the companion of the binary pulsar
and investigate the Shapiro time delate [13] due to the Kerr-Newmann
curvature produced by such a black hole. Instead of focusing the attention
on the black hole curvature we think about the black hole gravitational
waves as being the responsible of the timing effect on the binary pulsars.
Our work may be useful, among other things, to distinguish these two
possibilities. Actually, our formulation is so general that the companion of
the binary pulsar can be any other source of gravitational waves such as
supernovae or vibrating neutron stars.

Also, this work may be of special interest in connection with other related
works. Recently, Wang \textit{et al.} [14] made the proposal to construct a
50m radio telescope to measure pulsar timings. The Wang \textit{et al.}
considerations are based on the idea proposed some time ago by Detweiler
[15] who showed that measurements of signal arrival time from a pulsar may
be used to search for stochastic gravitational wave background (SGWB). The
main strategy of these authors to detect of the SGWB is to consider a number
of pulsars separated at different parts in the sky. It is clear then that
our RTDE formulation may be useful for the project of these authors. Also
recently, Kessari \textit{et al.} [16] (see also Refs. [17]and [18])
extended the idea of Ref. [7] to include electromagnetic waves. Since the
polarization vector of an electromagnetic wave can be described by the spin
tensor of a massless top, it seems reasonable to think that the RTDE
approach may also be of special interest in this direction.

Finally, the RTDE may also have an interesting application in connection
with the so called inflationary gravitational waves (see [19] and references
there in). As it is known, the polarization of the cosmic microwave
radiation [20] may solve the problem of detecting the gravitational waves
produced during the inflationary scenario. Just before the universe became
transparent to radiation, the plasma motion caused by the gravitational
waves may have different sources. In particular, the effect predicted by the
RTDE may be of particular interest in this scenario.

Moreover, inflationary gravitational waves are predicted by higher
dimensional theories such as string/M theories [21]-[24] and supergravity.
An attractive scenario with extra dimensions is the brane worlds cosmology.
In this case, our 3+1 dimensional spacetime is the dynamical 3d brane
embedded in the higher dimensional space. The information about the
bulk/brane geometry can be obtained through the gravitational waves. All
particles in the standard model are confined to the brane and they cannot
move in the higher dimensional space, with the sole exception of gravity.
Therefore, if we identify the internal angular momentum of the top with the
spin tensor of a fundamental particle, then the RTDE provides an interaction
between brane world gravitational waves and the spin of the particles in the
plasma contained in the brane. Thus, the RTDE system may be useful to get
information through gravitational excitations.

The plan of this work is as follows. In section II, we briefly review one of
the possible mechanisms to obtain the GDE and in section III we apply
similar techniques to obtain the RTDE formulation. In section IV, we explain
how the RTDE can be applied to the detection of gravitational waves.
Finally, in section V, we make some final comments.

\bigskip \smallskip \bigskip\ 

\noindent \textbf{II. GEODESIC DEVIATION EQUATION}

\bigskip

Several methods can be used to obtain the GDE. Some of them are, in fact,
quite brief. For our purpose, however, it turns out to be more convenient to
follow the one in reference [25], emphasizing that our computations are more
specific than those of [25].

Consider a point particle whose trajectory is described by the coordinates $%
x^{\mu}(\tau),$ where $\tau$ is the proper time parameter. The geodesic
equation is

\begin{equation}
\frac{d^{2}x^{\mu }}{d\tau ^{2}}+\Gamma _{\alpha \beta }^{\mu }(x)\frac{%
dx^{\alpha }}{d\tau }\frac{dx^{\beta }}{d\tau }=0.  \label{1}
\end{equation}
Here, $\Gamma _{\alpha \beta }^{\mu }(x)$ stands for the Christoffel symbols.

Equation (1) can be written in a more compact form as

\begin{equation}
\frac{D^{2}x^{\mu }}{D\tau ^{2}}=0,  \label{2}
\end{equation}
where $\frac{D}{D\tau }$ denotes covariant derivative with respect to $\tau $
and $\frac{D^{2}x^{\mu }}{D\tau ^{2}}=\frac{D}{D\tau }(\frac{Dx^{\mu }}{%
D\tau })$. Note that, since the coordinates $x^{\mu }$ are scalars fields,
one has $\frac{Dx^{\mu }}{D\tau }=\frac{dx^{\mu }}{d\tau }.$

A nearby point particle must also satisfy the geodesic equation. If we use
the coordinates $x^{\prime \mu }(\tau )$ to describe the position of such a
nearby point particle we have

\begin{equation}
\frac{d^{2}x^{\prime\mu}}{d\tau^{2}}+\Gamma_{\alpha\beta}^{\prime\mu
}(x^{\prime})\frac{dx^{\prime\alpha}}{d\tau}\frac{dx^{\prime\beta}}{d\tau}=0.
\label{3}
\end{equation}
By nearby we mean that the coordinates $x^{\prime\mu}(\tau)$ can be written
as

\begin{equation}
x^{\prime \mu }=x^{\mu }+\xi ^{\mu }(x),  \label{4}
\end{equation}
with $\shortmid \xi ^{\mu }\shortmid $ being a very small quantity.

To first order in $\xi ^{\mu }$ we have

\begin{equation}
\Gamma _{\alpha \beta }^{\prime \mu }(x+\xi )=\Gamma _{\alpha \beta }^{\mu
}(x)+\Gamma _{\alpha \beta }^{\mu },_{\lambda }\xi ^{\lambda },  \label{5}
\end{equation}
with $\Gamma _{\alpha \beta }^{\mu },_{\lambda }=\frac{\partial \Gamma
_{\alpha \beta }^{\mu }}{\partial x^{\lambda }}$. Thus, using (4) and (5) we
find that equation (3) becomes

\begin{equation}
\frac{d^{2}x^{\mu }}{d\tau ^{2}}+\frac{d^{2}\xi ^{\mu }}{d\tau ^{2}}+\Gamma
_{\alpha \beta }^{\mu }\frac{dx^{\alpha }}{d\tau }\frac{dx^{\beta }}{d\tau }%
+2\Gamma _{\alpha \beta }^{\mu }\frac{dx^{\alpha }}{d\tau }\frac{d\xi
^{\beta }}{d\tau }+\Gamma _{\alpha \beta }^{\mu },_{\lambda }\xi ^{\lambda }%
\frac{dx^{\alpha }}{d\tau }\frac{dx^{\beta }}{d\tau }=0.  \label{6}
\end{equation}
By virtue of (1), we see that the first and the third term of (6) can be
dropped. Hence, the expression (6) is reduced to 
\begin{equation}
\frac{d^{2}\xi ^{\mu }}{d\tau ^{2}}+2\Gamma _{\alpha \beta }^{\mu }\frac{%
dx^{\alpha }}{d\tau }\frac{d\xi ^{\beta }}{d\tau }+\Gamma _{\alpha \beta
}^{\mu },_{\lambda }\xi ^{\lambda }\frac{dx^{\alpha }}{d\tau }\frac{%
dx^{\beta }}{d\tau }=0.  \label{7}
\end{equation}
Our goal is now to write this equation in a covariant form. For this purpose
we first write (7) in the form

\begin{equation}
\frac{d^{2}\xi ^{\mu }}{d\tau ^{2}}+2\Gamma _{\alpha \beta }^{\mu }\frac{%
dx^{\alpha }}{d\tau }\frac{d\xi ^{\beta }}{d\tau }=-\Gamma _{\alpha \beta
}^{\mu },_{\lambda }\xi ^{\lambda }\frac{dx^{\alpha }}{d\tau }\frac{%
dx^{\beta }}{d\tau }.  \label{8}
\end{equation}
Now, by adding to both sides of (8) the expression

\begin{equation}
\Gamma _{\alpha \lambda }^{\mu },_{\beta }\frac{dx^{\alpha }}{d\tau }\frac{%
dx^{\beta }}{d\tau }\xi ^{\lambda }+\Gamma _{\sigma \beta }^{\mu }\Gamma
_{\alpha \lambda }^{\sigma }\frac{dx^{\alpha }}{d\tau }\frac{dx^{\beta }}{%
d\tau }\xi ^{\lambda }-\Gamma _{\sigma \lambda }^{\mu }\Gamma _{\alpha \beta
}^{\sigma }\frac{dx^{\alpha }}{d\tau }\frac{dx^{\beta }}{d\tau }\xi
^{\lambda }.  \label{9}
\end{equation}
we observe that (8) becomes

\begin{equation}
\begin{array}{c}
\frac{d^{2}\xi ^{\mu }}{d\tau ^{2}}+2\Gamma _{\alpha \beta }^{\mu }\frac{%
dx^{\alpha }}{d\tau }\frac{d\xi ^{\beta }}{d\tau }+\Gamma _{\alpha \lambda
}^{\mu },_{\beta }\frac{dx^{\alpha }}{d\tau }\frac{dx^{\beta }}{d\tau }\xi
^{\lambda } \\ 
\\ 
+\Gamma _{\sigma \beta }^{\mu }\Gamma _{\alpha \lambda }^{\sigma }\frac{%
dx^{\alpha }}{d\tau }\frac{dx^{\beta }}{d\tau }\xi ^{\lambda }-\Gamma
_{\sigma \lambda }^{\mu }\Gamma _{\alpha \beta }^{\sigma }\frac{dx^{\alpha }%
}{d\tau }\frac{dx^{\beta }}{d\tau }\xi ^{\lambda }=-R_{\alpha \lambda \beta
}^{\mu }\frac{dx^{\alpha }}{d\tau }\xi ^{\lambda }\frac{dx^{\beta }}{d\tau },
\end{array}
\label{10}
\end{equation}
where $R_{\alpha \lambda \beta }^{\mu }$ is the usual curvature Riemann
tensor,

\begin{equation}
R_{\alpha \lambda \beta }^{\mu }=\Gamma _{\alpha \beta }^{\mu },_{\lambda
}-\Gamma _{\alpha \lambda }^{\mu },_{\beta }+\Gamma _{\sigma \lambda }^{\mu
}\Gamma _{\alpha \beta }^{\sigma }-\Gamma _{\sigma \beta }^{\mu }\Gamma
_{\alpha \lambda }^{\sigma }.  \label{11}
\end{equation}

By considering (1), we note that the fifth term of the left hand side of
(10) can be written as

\begin{equation}
-\Gamma _{\sigma \lambda }^{\mu }\Gamma _{\alpha \beta }^{\sigma }\frac{%
dx^{\alpha }}{d\tau }\frac{dx^{\beta }}{d\tau }\xi ^{\lambda }=\Gamma
_{\sigma \lambda }^{\mu }\frac{d^{2}x^{\sigma }}{d\tau ^{2}}\xi ^{\lambda }.
\label{12}
\end{equation}
Substituting this result into (10), it is not difficult to see that (10) can
be written in the covariant form

\begin{equation}
\frac{D^{2}\xi ^{\mu }}{D\tau ^{2}}=-R_{\alpha \lambda \beta }^{\mu }\frac{%
dx^{\alpha }}{d\tau }\xi ^{\lambda }\frac{dx^{\beta }}{d\tau },  \label{13}
\end{equation}
which is, of course, the famous geodesic deviation equation (GDE) for a pair
of nearby freely falling particles in a gravitational field background.

\bigskip\bigskip\bigskip\bigskip\ 

\noindent \textbf{III. RELATIVISTIC TOP DEVIATION EQUATION}

\bigskip

The equations of motion of a relativistic top moving in a gravitational
field background are

\begin{equation}
\frac{D^{2}x^{\mu}}{D\tau^{2}}=-\frac{1}{2}R_{\alpha\lambda\beta}^{\mu}\frac{%
dx^{\alpha}}{d\tau}S^{\lambda\beta},  \label{14}
\end{equation}
and

\begin{equation}
\frac{DS^{\mu \nu }}{D\tau }=0.  \label{15}
\end{equation}
Here $S^{\mu \nu }=-S^{\nu \mu }$ is the internal angular momentum (or the
spin tensor) per unit mass of the top satisfying the Pirani constraint [26] $%
S^{\mu \nu }\frac{dx_{\nu }}{d\tau }=0$. It is worth mentioning that the
formulae (14) and (15) can be derived from a number of different methods [8]
(see also Refs. [10], and references there in). Perhaps two of the most
interesting are the Lagrangian formulation due to Rietdijk--Holten [9] and
Galvao and Teitelboim [27] (see Refs. [28] and [29] for early works), and
Hojman [30]. In the Rietdijk-Holten-Galvao-Teitelboim (RHGT) approach the
spinning top is described by the variables $x^{\mu }(\tau )$ and $\theta
^{\mu }(\tau ),$ where the $\theta ^{\mu }(\tau )$ are anticommuting
variables (Grassmann coordinates), while in the Hojman formalism the
rotation of the top is described by the four vectors $e_{(\alpha )}^{\mu
}(\tau ).$ Specifically, in the case of RHGT the spin tensor $S^{\mu \nu }$
is given by $S^{\mu \nu }=i\theta ^{\mu }\theta ^{\nu }$, while in the
Hojman's approach $S^{\mu \nu }$ is the canonical momentum associated to the
angular velocity $\sigma ^{\mu \nu }=e^{\mu (\alpha )}\frac{D}{D\tau }%
e_{(\alpha )}^{\nu }$. In both cases the corresponding Lagrangian is taken
to be a Poincar\'{e} invariant. One of the advantages of the Lagrangian
formulation is that it allows the use of a variational principle to insure
the consistence of the equations of motion. However, the main motivation to
develop the Lagrangian formalism for the top is the desire to use the Dirac%
\'{}%
s constraint Hamiltonian method to quantize the system. \ An important
aspect of this construction is that it is more convenient to consider the
Tulczyjew constraints [31] $S^{\mu \nu }P_{\nu }=0,$ with $P_{\mu }$ the
linear momentum of the top, rather than the Pirani constraint. As a result
the linear momentum $P_{\mu }$ turns out to be non parallel to the velocity $%
u^{\mu }=\frac{dx^{\mu }}{d\tau }$. However this difference is very slight
and gauge dependent. Therefore, for practical purposes the Tulczyjew
constraint and the Pirani constraint are the same and the reduced equations
of motion look like (14) and (15). A final observation is that (14) can be
understood as the analogue of the geodesic equation (1) and in fact it
reduces to (1) when the spin tensor $S^{\mu \nu }$ vanishes.

After comparing equations (13) and (14) we observe a great similarity. But
in fact they are very different in the sense that while equation (13) refers
to a pair of nearby point particles, (14) is associated with just one
physical system: a relativistic top. Nevertheless, this similarity was used
as an inspiration to propose that just as (13) is used to detect
gravitational waves, equation (14) can be used for the same purpose. In
order to better understand the real differences between the two point
particles system and the relativistic top it is necessary to derive the
analogue of (13) for a pair of nearby relativistic tops. For this purpose
let us closely follow the method of section II, but now using the
relativistic top equation of motion (14) instead of the formula (1).

A nearby top must satisfy the corresponding equation of motion

\begin{equation}
\frac{d^{2}x^{\prime\mu}}{d\tau^{2}}+\Gamma_{\alpha\beta}^{\prime\mu
}(x^{\prime})\frac{dx^{\prime\alpha}}{d\tau}\frac{dx^{\prime\beta}}{d\tau }=-%
\frac{1}{2}R_{\alpha\lambda\beta}^{\prime\mu}(x^{\prime})\frac
{dx^{\prime\alpha}}{d\tau}S^{\prime\lambda\beta}.  \label{16}
\end{equation}
Consider now a perturbation of the form

\begin{equation}
x^{\prime\mu}=x^{\mu}+\xi^{\mu}(x).  \label{17}
\end{equation}
and

\begin{equation}
S^{\prime\mu\nu}=S^{\mu\nu}+S^{\mu\nu},_{\alpha}\xi^{\alpha}(x).  \label{18}
\end{equation}

We are interested in developing the formula

\begin{equation}
\begin{array}{c}
\frac{d^{2}(x^{\mu }+\xi ^{\mu })}{d\tau ^{2}}+(\Gamma _{\alpha \beta }^{\mu
}(x)+\Gamma _{\alpha \beta }^{\mu },_{\lambda }\xi ^{\lambda })\frac{%
d(x^{\alpha }+\xi ^{\alpha })}{d\tau }\frac{d(x^{\beta }+\xi ^{\beta })}{%
d\tau } \\ 
\\ 
=-\frac{1}{2}(R_{\alpha \lambda \beta }^{\mu }(x)+R_{\alpha \lambda \beta
}^{\mu },_{\sigma }\xi ^{\sigma })(\frac{d(x^{\alpha }+\xi ^{\alpha })}{%
d\tau })(S^{\lambda \beta }+S^{\lambda \beta },_{\gamma }\xi ^{\gamma })
\end{array}
\label{19}
\end{equation}
to first order in $\xi ^{\mu }.$ We find

\begin{equation}
\begin{array}{c}
\frac{d^{2}x^{\mu }}{d\tau ^{2}}+\frac{d^{2}\xi ^{\mu }}{d\tau ^{2}}+\Gamma
_{\alpha \beta }^{\mu }\frac{dx^{\alpha }}{d\tau }\frac{dx^{\beta }}{d\tau }%
+2\Gamma _{\alpha \beta }^{\mu }\frac{dx^{\alpha }}{d\tau }\frac{d\xi
^{\beta }}{d\tau }+\Gamma _{\alpha \beta }^{\mu },_{\lambda }\xi ^{\lambda }%
\frac{dx^{\alpha }}{d\tau }\frac{dx^{\beta }}{d\tau } \\ 
\\ 
=-\frac{1}{2}[R_{\alpha \lambda \beta }^{\mu }\frac{dx^{\alpha }}{d\tau }%
S^{\lambda \beta }+R_{\alpha \lambda \beta }^{\mu }\frac{d\xi ^{\alpha }}{%
d\tau }S^{\lambda \beta }+R_{\alpha \lambda \beta }^{\mu }\frac{dx^{\alpha }%
}{d\tau }S^{\lambda \beta },_{\gamma }\xi ^{\gamma }+R_{\alpha \lambda \beta
}^{\mu },_{\sigma }\xi ^{\sigma }\frac{dx^{\alpha }}{d\tau }S^{\lambda \beta
}].
\end{array}
\label{20}
\end{equation}
By using (14), this formula is simplified to

\begin{equation}
\begin{array}{c}
\frac{d^{2}\xi ^{\mu }}{d\tau ^{2}}+2\Gamma _{\alpha \beta }^{\mu }\frac{%
dx^{\alpha }}{d\tau }\frac{d\xi ^{\beta }}{d\tau }+\Gamma _{\alpha \beta
}^{\mu },_{\lambda }\xi ^{\lambda }\frac{dx^{\alpha }}{d\tau }\frac{%
dx^{\beta }}{d\tau } \\ 
\\ 
=-\frac{1}{2}[R_{\alpha \lambda \beta }^{\mu }\frac{d\xi ^{\alpha }}{d\tau }%
S^{\lambda \beta }+R_{\alpha \lambda \beta }^{\mu }\frac{dx^{\alpha }}{d\tau 
}S^{\lambda \beta },_{\gamma }\xi ^{\gamma }+R_{\alpha \lambda \beta }^{\mu
},_{\sigma }\xi ^{\sigma }\frac{dx^{\alpha }}{d\tau }S^{\lambda \beta }].
\end{array}
\label{21}
\end{equation}
It is more convenient to write this equation as

\begin{equation}
\begin{array}{c}
\frac{d^{2}\xi ^{\mu }}{d\tau ^{2}}+2\Gamma _{\alpha \beta }^{\mu }\frac{%
dx^{\alpha }}{d\tau }\frac{d\xi ^{\beta }}{d\tau }=-\Gamma _{\alpha \beta
}^{\mu },_{\lambda }\xi ^{\lambda }\frac{dx^{\alpha }}{d\tau }\frac{%
dx^{\beta }}{d\tau } \\ 
\\ 
-\frac{1}{2}[R_{\alpha \lambda \beta }^{\mu }\frac{d\xi ^{\alpha }}{d\tau }%
S^{\lambda \beta }+R_{\alpha \lambda \beta }^{\mu }\frac{dx^{\alpha }}{d\tau 
}S^{\lambda \beta },_{\gamma }\xi ^{\gamma }+R_{\alpha \lambda \beta }^{\mu
},_{\sigma }\xi ^{\sigma }\frac{dx^{\alpha }}{d\tau }S^{\lambda \beta }],
\end{array}
\label{22}
\end{equation}
The next step is to write this formula in a covariant form.

For this purpose, as in section II, we add to both sides of (22) the
expression

\begin{equation}
\Gamma_{\alpha\lambda}^{\mu},_{\beta}\frac{dx^{\alpha}}{d\tau}\frac{%
dx^{\beta }}{d\tau}\xi^{\lambda}+\Gamma_{\sigma\beta}^{\mu}\Gamma_{\alpha%
\lambda }^{\sigma}\frac{dx^{\alpha}}{d\tau}\frac{dx^{\beta}}{d\tau}%
\xi^{\lambda }-\Gamma_{\sigma\lambda}^{\mu}\Gamma_{\alpha\beta}^{\sigma}%
\frac{dx^{\alpha}}{d\tau}\frac{dx^{\beta}}{d\tau}\xi^{\lambda}.  \label{23}
\end{equation}
We get

\begin{equation}
\begin{array}{c}
\frac{d^{2}\xi ^{\mu }}{d\tau ^{2}}+2\Gamma _{\alpha \beta }^{\mu }\frac{%
dx^{\alpha }}{d\tau }\frac{d\xi ^{\beta }}{d\tau }+\Gamma _{\alpha \lambda
}^{\mu },_{\beta }\frac{dx^{\alpha }}{d\tau }\frac{dx^{\beta }}{d\tau }\xi
^{\lambda }+\Gamma _{\sigma \beta }^{\mu }\Gamma _{\alpha \lambda }^{\sigma }%
\frac{dx^{\alpha }}{d\tau }\frac{dx^{\beta }}{d\tau }\xi ^{\lambda }-\Gamma
_{\sigma \lambda }^{\mu }\Gamma _{\alpha \beta }^{\sigma }\frac{dx^{\alpha }%
}{d\tau }\frac{dx^{\beta }}{d\tau }\xi ^{\lambda } \\ 
\\ 
=-R_{\alpha \lambda \beta }^{\mu }\frac{dx^{\alpha }}{d\tau }\xi ^{\lambda }%
\frac{dx^{\beta }}{d\tau }-\frac{1}{2}[R_{\alpha \lambda \beta }^{\mu }\frac{%
d\xi ^{\alpha }}{d\tau }S^{\lambda \beta }+R_{\alpha \lambda \beta }^{\mu }%
\frac{dx^{\alpha }}{d\tau }S^{\lambda \beta },_{\gamma }\xi ^{\gamma
}+R_{\alpha \lambda \beta }^{\mu },_{\sigma }\xi ^{\sigma }\frac{dx^{\alpha }%
}{d\tau }S^{\lambda \beta }],
\end{array}
\label{24}
\end{equation}
where we have used the definition of the Riemann tensor $R_{\alpha \lambda
\beta }^{\mu }.$

In contrast with (12), using (14) we now find that the fifth term in the
left hand side of (24) leads to

\begin{equation}
-\Gamma _{\sigma \lambda }^{\mu }\Gamma _{\alpha \beta }^{\sigma }\frac{%
dx^{\alpha }}{d\tau }\frac{dx^{\beta }}{d\tau }\xi ^{\lambda }=\Gamma
_{\sigma \lambda }^{\mu }\frac{d^{2}x^{\sigma }}{d\tau ^{2}}\xi ^{\lambda }+%
\frac{1}{2}\Gamma _{\sigma \gamma }^{\mu }\xi ^{\gamma }R_{\alpha \lambda
\beta }^{\sigma }\frac{dx^{\alpha }}{d\tau }S^{\lambda \beta }.  \label{25}
\end{equation}
Substituting (25) into (24), we learn that (24) can be written as

\begin{equation}
\begin{array}{c}
\frac{D^{2}\xi^{\mu}}{D\tau^{2}}+\frac{1}{2}\Gamma_{\sigma\gamma}^{\mu}\xi^{%
\gamma}R_{\alpha\lambda\beta}^{\sigma}\frac{dx^{\alpha}}{d\tau}%
S^{\lambda\beta}=-R_{\alpha\lambda\beta}^{\mu}\frac{dx^{\alpha}}{d\tau}%
\xi^{\lambda}\frac{dx^{\beta}}{d\tau} \\ 
\\ 
-\frac{1}{2}[R_{\alpha\lambda\beta}^{\mu}\frac{d\xi^{\alpha}}{d\tau}%
S^{\lambda\beta}+R_{\alpha\lambda\beta}^{\mu}\frac{dx^{\alpha}}{d\tau }%
S^{\lambda\beta},_{\gamma}\xi^{\gamma}+R_{\alpha\lambda\beta}^{\mu},_{\sigma
}\xi^{\sigma}\frac{dx^{\alpha}}{d\tau}S^{\lambda\beta}].
\end{array}
\label{26}
\end{equation}

It is not difficult to show that

\begin{equation}
\begin{array}{c}
-\frac{1}{2}[R_{\alpha \lambda \beta }^{\mu }\frac{d\xi ^{\alpha }}{d\tau }%
S^{\lambda \beta }+R_{\alpha \lambda \beta }^{\mu }\frac{dx^{\alpha }}{d\tau 
}S^{\lambda \beta },_{\gamma }\xi ^{\gamma }+R_{\alpha \lambda \beta }^{\mu
},_{\sigma }\xi ^{\sigma }\frac{dx^{\alpha }}{d\tau }S^{\lambda \beta }] \\ 
\\ 
=-\frac{1}{2}[R_{\alpha \lambda \beta }^{\mu }\frac{D\xi ^{\alpha }}{D\tau }%
S^{\lambda \beta }+R_{\alpha \lambda \beta }^{\mu }\frac{dx^{\alpha }}{d\tau 
}S^{\lambda \beta };_{\gamma }\xi ^{\gamma }+R_{\alpha \lambda \beta }^{\mu
};_{\sigma }\xi ^{\sigma }\frac{dx^{\alpha }}{d\tau }S^{\lambda \beta }] \\ 
\\ 
+\frac{1}{2}\Gamma _{\sigma \gamma }^{\mu }\xi ^{\gamma }R_{\alpha \lambda
\beta }^{\sigma }\frac{dx^{\alpha }}{d\tau }S^{\lambda \beta }.
\end{array}
\label{27}
\end{equation}
Here, for any contravariant tensor $A^{\mu }$ we define the covariant
derivative as $A^{\mu };_{\sigma }=A^{\mu },_{\sigma }+\Gamma _{\sigma
\gamma }^{\mu }A^{\gamma }.$ Although the result (27) seems to be evident,
it is not a trivial one since for its obtainment many terms were cancelled.

We finally discover that using (27), the expression (26) becomes

\begin{equation}
\begin{array}{c}
\frac{D^{2}\xi ^{\mu }}{D\tau ^{2}}=-R_{\alpha \lambda \beta }^{\mu }\frac{%
dx^{\alpha }}{d\tau }\xi ^{\lambda }\frac{dx^{\beta }}{d\tau } \\ 
\\ 
-\frac{1}{2}[R_{\alpha \lambda \beta }^{\mu }\frac{D\xi ^{\alpha }}{D\tau }%
S^{\lambda \beta }+R_{\alpha \lambda \beta }^{\mu }\frac{dx^{\alpha }}{d\tau 
}S^{\lambda \beta };_{\gamma }\xi ^{\gamma }+R_{\alpha \lambda \beta }^{\mu
};_{\sigma }\xi ^{\sigma }\frac{dx^{\alpha }}{d\tau }S^{\lambda \beta }],
\end{array}
\label{28}
\end{equation}
which is the covariant form of the relativistic top deviation equation
(RTDE). Clearly, (28) reduces to (13) when the spin tensor $S^{\lambda \beta
}$ vanishes. Therefore, (28) is an extension of (13). (It is worth
mentioning that in Ref. [32] appears similar equations to those in (28).)
One of the attractive features of (28) is that the spin angular momentum $%
S^{\lambda \beta }$ of the top is coupled to gravity via the curvature
Riemann tensor $R_{\alpha \lambda \beta }^{\mu }$ and the gradient of this.
\ It seems reasonable to think that this characteristic can provide a better
description of the properties of the underlying curvature geometry. In
particular, we shall see in the next section that the RTDE may be used to
study different properties of a gravitational wave background.

\bigskip \smallskip \bigskip\ 

\noindent \textbf{IV. THE RELATIVISTIC TOP DEVIATION EQUATION AND
GRAVITATIONAL WAVES}

\bigskip

In this section we will investigate the consequences of equation (28) to the
case of gravitational waves. But for completeness we will start by reviewing
briefly how the formula (13) is used for this particular case.

Consider a gravitational wave in a flat background. For this case, the
metric $g_{\mu \nu }$ can be written as

\begin{equation}
g_{\mu \nu }=\eta _{\mu \nu }+h_{\mu \nu },  \label{29}
\end{equation}
where $\eta _{\mu \nu }$ is the Minkowski metric and $\mid h_{\mu \nu }\mid
\ll 1.$ \ In the transverse-traceless gauge

\begin{equation}
\begin{array}{c}
h_{0\mu }=0, \\ 
\\ 
h_{ij,j}=0, \\ 
\\ 
h_{\mu }^{\mu }=0,
\end{array}
\label{30}
\end{equation}
with the indices $i,j,...,etc$ running from $1$ to $3$. The Einstein
gravitational field equations imply that $h_{ij}$ satisfies the wave equation

\begin{equation}
\square ^{2}h_{ij}=0,  \label{31}
\end{equation}
where $\square ^{2}=\partial ^{\mu }\partial _{\mu }$ is the D'Alambertian.
In the gauge given in (30), the space-time components of the Riemann tensor $%
R_{i0j0}$ have the simple form

\begin{equation}
R_{i0j0}=-\frac{1}{2}h_{ij},_{00}.  \label{32}
\end{equation}

By considering (30), one discovers that $h_{ij}$ can be written as

\begin{equation}
h_{ij}=A_{+}e_{ij}^{+}+A_{\times }e_{ij}^{\times },  \label{33}
\end{equation}
where $A_{+}$ and $A_{\times }$ are two independent dimensionless amplitudes
and $e_{ij}^{+}$ and $e_{ij}^{\times }$ are polarization tensors. For a wave
traveling in the $z-$direction the only nonvanishing components of $e_{ij}$
are

\begin{equation}
\begin{array}{c}
e_{xx}^{+}=-e_{yy}^{+}, \\ 
\\ 
e_{xy}^{\times }=e_{yx}^{\times }
\end{array}
\label{34}
\end{equation}
and in this case $A_{+}$ and $A_{\times }$ turn out to be functions that
depend only on $t-z$.

In a proper reference frame we have $x^{0}=\tau ,$ $x^{i}=0,$ so that $\frac{%
dx^{0}}{d\tau }=1$ and $\frac{dx^{i}}{d\tau }=0.$ In this reference frame we
find that (13) becomes

\begin{equation}
\frac{d^{2}\xi ^{i}}{dt^{2}}=-R_{0j0}^{i}\xi ^{j},  \label{35}
\end{equation}
By using (32), we find that equation (35) can be rewritten as

\begin{equation}
\frac{d^{2}\xi ^{i}}{dt^{2}}=\frac{1}{2}h_{j}^{i},_{00}\xi ^{j},  \label{36}
\end{equation}
with $h_{j}^{i}=\delta ^{ik}h_{kj}$, and for a wave propagating in the $z-$%
direction, (36) implies

\begin{equation}
\frac{d^{2}\xi ^{z}}{dt^{2}}=0,  \label{37}
\end{equation}
as well as

\begin{equation}
\frac{d^{2}\xi ^{a}}{dt^{2}}=\frac{1}{2}h_{b}^{a},_{00}\xi ^{b},  \label{38}
\end{equation}
where now, the latin indices $a,b,..,etc$ run from $1$ to $2$. The formula
(38) tells us that only separations between two nearby point particles in
the transverse direction are meaningful.

Let us now address the problem at hand, namely, we are interested in
applying a similar method as the one above to the case of a system with two
nearby relativistic tops. For this purpose let us consider the formula (28)
in a proper reference frame. We have

\begin{equation}
\frac{d^{2}\xi ^{\mu }}{dt^{2}}=-R_{0k0}^{\mu }\xi ^{k}-\frac{1}{2}[%
R_{\alpha kl}^{\mu }\frac{d\xi ^{\alpha }}{d\tau }S^{kl}+R_{0kl}^{\mu
}S^{kl},_{\gamma }\xi ^{\gamma }+R_{0kl}^{\mu },_{\sigma }\xi ^{\sigma
}S^{kl}],  \label{39}
\end{equation}
where we used the fact that $S^{0\mu }=0$ due to the Twlczyjew-Pirani
constraint $S^{\mu \nu }\frac{dx_{\nu }}{d\tau }=0$.

Using the symmetries of the Riemann curvature tensor we find that the time
component of (39) is

\begin{equation}
\frac{d^{2}\xi^{0}}{dt^{2}}=-\frac{1}{2}R_{jkl}^{0}\frac{d\xi^{j}}{dt}S^{kl},
\label{40}
\end{equation}
while the space components become

\begin{equation}
\begin{array}{c}
\frac{d^{2}\xi^{i}}{dt^{2}}=-R_{0k0}^{i}\xi^{k}-\frac{1}{2}[R_{0kl}^{i}\frac{%
d\xi^{0}}{dt}S^{kl}+R_{0kl}^{i}S^{kl},_{0}\xi^{0}+R_{0kl}^{i},_{0}%
\xi^{0}S^{kl}] \\ 
\\ 
-\frac{1}{2}[R_{jkl}^{i}\frac{d\xi^{j}}{dt}S^{kl}+R_{0kl}^{i}S^{kl},_{j}%
\xi^{j}+R_{0kl}^{i},_{j}\xi^{j}S^{kl}].
\end{array}
\label{41}
\end{equation}

We shall now apply (40) and (41) for the particular case of a gravitational
plane wave propagating in the $z-$direction. It is convenient to write the
indices $i,j...etc$ as ($a,z).$ With this notation (40) can be written as

\begin{equation}
\frac{d^{2}\xi ^{0}}{dt^{2}}=-R_{azb}^{0}\frac{d\xi ^{a}}{dt}S^{zb},
\label{42}
\end{equation}
where we used the fact that the only nonvanishing components of the Riemann
curvature tensor are

\begin{equation}
R_{zazb}=R_{0a0b}=-R_{0azb}=-\frac{1}{2}h_{ab,00}.  \label{43}
\end{equation}
While (41) yields

\begin{equation}
\begin{array}{c}
\frac{d^{2}\xi ^{i}}{dt^{2}}=-R_{0b0}^{i}\xi ^{b}-[R_{0zb}^{i}\frac{d\xi ^{0}%
}{dt}S^{zb}+R_{0zb}^{i}S^{zb},_{0}\xi ^{0}+R_{0zb}^{i},_{0}\xi ^{0}S^{zb}]
\\ 
\\ 
-\frac{1}{2}[R_{0ab}^{i}\frac{d\xi ^{0}}{dt}S^{ab}+R_{0ab}^{i}S^{ab},_{0}\xi
^{0}+R_{0ab}^{i},_{0}\xi ^{0}S^{ab}] \\ 
\\ 
-[R_{zbz}^{i}\frac{d\xi ^{z}}{dt}S^{bz}+R_{0bz}^{i}S^{bz},_{z}\xi
^{z}+R_{0bz}^{i},_{z}\xi ^{z}S^{bz}] \\ 
\\ 
-\frac{1}{2}[R_{zab}^{i}\frac{d\xi ^{z}}{dt}S^{ab}+R_{0ab}^{i}S^{ab},_{z}\xi
^{z}+R_{0ab}^{i},_{z}\xi ^{z}S^{ab}] \\ 
\\ 
-[R_{abz}^{i}\frac{d\xi ^{a}}{dt}S^{bz}+R_{0bz}^{i}S^{bz},_{a}\xi
^{a}+R_{0bz}^{i},_{a}\xi ^{a}S^{bz}] \\ 
\\ 
-\frac{1}{2}[R_{abc}^{i}\frac{d\xi ^{a}}{dt}S^{bc}+R_{0bc}^{i}S^{bc},_{a}\xi
^{a}+R_{0bc}^{i},_{a}\xi ^{a}S^{bc}]
\end{array}
\label{44}
\end{equation}

The$\ z$ component of (44) is

\begin{equation}
\frac{d^{2}\xi ^{z}}{dt^{2}}=-R_{azb}^{z}\frac{d\xi ^{a}}{dt}S^{zb},
\label{45}
\end{equation}
where we used (43), and the $x$ and $y$ components are

\begin{equation}
\begin{array}{c}
\frac{d^{2}\xi ^{a}}{dt^{2}}=-R_{0b0}^{a}\xi ^{b}-[R_{0bz}^{a}\frac{d\xi ^{0}%
}{dt}S^{bz}+R_{0bz}^{a}S^{bz},_{0}\xi ^{0}+R_{0bz}^{a},_{0}\xi ^{0}S^{bz}]
\\ 
\\ 
-[R_{zbz}^{a}\frac{d\xi ^{z}}{dt}S^{bz}+R_{0bz}^{a}S^{bz},_{z}\xi
^{z}+R_{0bz}^{a},_{z}\xi ^{z}S^{bz}]-R_{0bz}^{a}S^{bz},_{a}\xi ^{a}.
\end{array}
\label{46}
\end{equation}
Note that the $S^{ab}$ component of the spin angular momentum does not
appear in (42), (45) and (46) and that only the $S^{zb}$ component remains$.$
To better understand this let us define

\begin{equation}
S^{i}=\frac{1}{2}\varepsilon ^{ijk}S_{jk},  \label{47}
\end{equation}
where $\varepsilon ^{ijk}$ is the Levi-Civita tensor with $\varepsilon
^{xyz}=1.$ From (47) we see that the $z$ component of the intrinsic angular
momentum does not play any role in equations (42), (45) and (46). This means
that no effect is expected when the top is oriented along the direction of
propagation of the gravitational wave .

The second interesting observation is that if $S^{zb}$ is nonvanishing then $%
\frac{d^{2}\xi ^{0}}{dt^{2}}\neq 0$ and $\frac{d^{2}\xi ^{z}}{dt^{2}}\neq 0,$
in contrast with the case of a pair of nearby point particles in which both
of these terms vanish. The third important observation is that the $\frac{%
d^{2}\xi ^{a}}{dt^{2}}$ equation contains a large number of terms in
addition to the usual one $R_{0b0}^{a}\xi ^{b}.$ Clearly, this means that
the solution of (46) will not be as simple as in the case of nearby point
particles.

Let us focus on the terms in (46) not involving derivatives of $S^{zb}$ and
the Riemann tensor, which presumably represent small order corrections. In
this case (46) is reduced to

\begin{equation}
\frac{d^{2}\xi ^{a}}{dt^{2}}=-R_{0b0}^{a}\xi ^{b}-R_{0bz}^{a}\frac{d\xi ^{0}%
}{dt}S^{bz}-R_{zbz}^{a}\frac{d\xi ^{z}}{dt}S^{bz}.  \label{48}
\end{equation}

By means of (43), we find that (42), (45) and (48) become

\begin{equation}
\frac{d^{2}\xi^{0}}{dt^{2}}=-\frac{1}{2}h_{ab,00}\frac{d\xi^{a}}{dt}S^{zb},
\label{49}
\end{equation}

\begin{equation}
\frac{d^{2}\xi ^{z}}{dt^{2}}=\frac{1}{2}h_{ab,00}\frac{d\xi ^{a}}{dt}S^{zb}
\label{50}
\end{equation}

and

\begin{equation}
\frac{d^{2}\xi ^{a}}{dt^{2}}=\frac{1}{2}h_{b}^{a},_{00}\xi ^{b}-\frac{1}{2}%
h_{b}^{a},_{00}\frac{d\xi ^{0}}{dt}S^{bz}+\frac{1}{2}h_{b}^{a},_{00}\frac{%
d\xi ^{z}}{dt}S^{bz},  \label{51}
\end{equation}
respectively. From (49) and (50) we discover that we can set $\xi ^{0}=-\xi
^{z}+cte.$ Therefore (51) becomes

\begin{equation}
\frac{d^{2}\xi ^{a}}{dt^{2}}=\frac{1}{2}h_{b}^{a},_{00}\xi
^{b}-h_{b}^{a},_{00}\frac{d\xi ^{z}}{dt}S^{bz}.  \label{52}
\end{equation}
Observe that if at an initial time the spin of the top $S^{bz}$ has an
orientation such that

\begin{equation}
\xi^{b}-2\frac{d\xi^{z}}{dt}S^{bz}=0,  \label{53}
\end{equation}
then there is not transverse motion $\xi^{a}=const.$ and therefore (50)
admits the solution

\begin{equation}
\xi ^{z}=const.  \label{54}
\end{equation}
What this means is that if the spin of the top is oriented along the vector
separation $\xi ^{a}$ of the two tops, then the gravitational wave does not
produce any effect on the system. This appears to be a new interesting
result since in the ordinary case of GDE the wave is always transverse in
its physical effects.

Let us now look for a solution of (52) of the form

\begin{equation}
\xi ^{a}=\xi _{0}^{a}+\frac{1}{2}h_{b}^{a}\xi _{0}^{b}-h_{b}^{a}(\frac{d\xi
^{z}}{dt}\mid _{0})S^{bz}.  \label{55}
\end{equation}
where $\xi _{0}^{a}=const$. We observe that $\frac{d\xi ^{a}}{dt}=\frac{1}{2}%
h_{b}^{a},_{0}\xi _{0}^{b}-h_{b}^{a},_{0}(\frac{d\xi ^{z}}{dt}\mid
_{0})S^{bz}$. Substituting (55) into (50) we find that to first order in $h$
(50) becomes

\begin{equation}
\frac{d^{2}\xi ^{z}}{dt^{2}}\approx 0,  \label{56}
\end{equation}
Therefore if initially $\frac{d\xi ^{z}}{dt}\mid _{0}=0$ to first order of
approximation the solution (55) reduces to the ordinary case of a pair
nearby point particles.

\bigskip \smallskip\ 

\noindent \textbf{V. COMMENTS}

\bigskip

We have been pursuing the possibility of using relativistic tops as
detectors of gravitational waves. In two previous works (Refs. [6] and [7])
an isolated top in a gravitational field was considered, making difficult to
compare the results with the case of DGE. In order to overcome this
difficulty and to gain further progress towards our goal, in this article we
have derived the RTDE equation for a pair of nearby tops. We have shown that
the RTDE reduces to the GDE when the spin tensor vanishes.

By considering a plane gravitational wave we find the solution of the RTDE
for two simple cases. In the first case we discover that if the internal
angular momentum of the top is oriented along the vector separation of the
two tops, the gravitational wave does not produce any effect on the physical
system. In the second more general case, we find that the nearby top will
have an effect different from the case of GDE only to a second order of
approximation in the perturbation $h_{ab}$. At first sight, it could seem
that these two cases show that even though the RTDE formulation it may be
theoretically interesting it does not offer a promising route for
experiments. However, rough estimates can show that this is not the case.

Let us write (55) in the form

\begin{equation}
\xi ^{a}=\xi _{0}^{a}+\frac{1}{2}h_{b}^{a}\xi _{0}^{b}+\eta ^{a},  \label{57}
\end{equation}
where

\begin{equation}
\eta ^{a}=h_{b}^{a}(\frac{d\xi ^{z}}{dt}\mid _{0})S^{bz}.  \label{58}
\end{equation}
Here, we are interested in making a rough estimate for $\eta ^{a}$. The
result $\xi ^{0}=-\xi ^{z}+const.$ allows to set $(\frac{d\xi ^{z}}{dt}\mid
_{0})\sim 1.$ Thus, (58) is reduced to

\begin{equation}
\eta ^{a}\sim h_{b}^{a}\frac{S^{bz}}{m_{0}c},  \label{59}
\end{equation}
where, in order to have the correct units, we restored the constants $m_{0}$
and $c.$ (Here, $m_{0}$ is the mass of the top and $c$ denotes the light
velocity.) In an order of magnitude, the expression (59) can be written in
the form

\begin{equation}
\eta \sim h\frac{S}{m_{0}c}.  \label{60}
\end{equation}
Roughly we can write $S\sim m_{0}r^{2}\omega $, so (60) becomes

\begin{equation}
\eta \sim h\frac{r^{2}\omega }{c}.  \label{61}
\end{equation}

For typical milli second (ms) pulsars $\omega \sim 10^{3}Hz$ and $r\sim
10^{6}cm$. Therefore, $\frac{r^{2}\omega }{c}\sim 10^{15}cm$ and from (61)
we find that

\begin{equation}
\eta \sim h(10^{5}cm).  \label{62}
\end{equation}

If we consider the best expected sensitivities of the wave amplitude $h$
over the earth, roughly $h\sim 10^{-18}$, we find that $\eta \sim
10^{-13}cm. $ This value for $\eta $ is, of course, too small for current
research detections. However, this estimate is based on the value $h\sim
10^{-18},$ which corresponds to gravitational radiation sources at distances
of Mpc from the Earth, but it may not necessarily correspond to some places
in the interstellar medium where the arrival of gravitational waves can have
a stronger effect. For instance, a pulsar may be close enough to a strong
source of gravitational waves companion such as another pulsar, black hole
or supernova. If this is the case, from the formula

\begin{equation}
h_{jk}=\frac{2}{r}\frac{G}{c^{2}}\frac{d^{2}Q_{jk}}{dt^{2}},  \label{63}
\end{equation}
where $Q_{jk}$ is the mass quadrupole moment, one finds that the expected
value for $h$ should be increased by several orders of magnitude. For
example, it is known that some globular clusters may be considered as
kitchens for making several ms pulsars. In such a scenario, from the rough
estimate of (63)

\begin{equation}
h\sim \frac{r_{sch}}{r}\frac{r_{sch}}{R},  \label{64}
\end{equation}
where $r_{sch}$ is the Schwarzschild radius and $R$ is the radius of the
source system, one finds that the value of $h$ could be, for instance, of
the order $10^{-1}$, which leads to a sensitivity of $\eta \sim 10^{4}cm.$
This result appears to be large enough to be detected from the arrival radio
signal coming from a pulsar.

A possible interesting extension of this work is to apply a similar method
to the one used to obtain the RTDE in the case of the nongeodesic equations
of motion of spinning particles in a teleparallel gravitational background
(GETGB) [33]-[34]. The GETGB equations are an extension of the RTEM in the
sense that they include torsion interactions in addition to the gravity spin
interaction. Recently, Garcia [35] has revisited the motion of spinless
particles in a teleparallel gravitational waves background and proposed an
experimental mechanism to detect the torsion via the GETGB model. The
complete picture should be to detect both gravitational waves and torsional
waves and therefore it may be interesting for further research to find the
non-deviation geodesic equations associated to the GETGB model.

Finally, we would like to comment about the possibility of using the RTDE in
connection with the inflationary gravitational waves scenario and brane new
world cosmology. In the first case, just before the universe became
transparent, the gravitational waves, produced during inflation, interacted
with the plasma producing polarization patterns of the cosmic microwave
background (CMB). This kind of phenomena is especially interesting since
recently some observations have shown the variation of the polarization
pattern of the CMB. The RTDE model can be interesting in this direction if
spin tensor $S^{\lambda \beta }$ is identified as the fermionic spin ($%
S^{\mu \nu }=i\theta ^{\nu }\theta ^{\mu }$) of elementary particles.
According to the RTDE, a gravitational wave should cause two kinds of
motions on the constituent fermionic particles of such a plasma. The first
one is the motion of the particles caused by the forces due to GDE and the
second one is the motion on the plasma caused by the spin-gravity
interaction. Therefore, one should expect that the spin-gravity interaction
may also leave a ``print'' in the polarization pattern of the CMB.

In the case of the world brane universes, only gravity can move in the extra
dimensions, all the matter and other forces are confined to the branes. Only
gravitational waves (or gravitons) travel from brane to brane carrying some
energy information away from the branes. Presumably, such gravitational
waves affect objects held together by gravity, such as stars and galaxies,
by distances shorter than millimeters. But, according to our above rough
estimate such short distance changes are also predicted by the RTDE model.

\bigskip

\textbf{Acknowledgment: }J. A. Nieto would like to thank the DIFUS and the
Astronomy Area for their hospitality.

\end{document}